\documentclass[aps, prl, twocolumn, superscriptaddress]{revtex4-1}
\usepackage{amsmath}
\usepackage{amssymb}
\usepackage[pdfusetitle, bookmarks = true,
            bookmarksnumbered = false,
            bookmarksopen, bookmarksopenlevel = 1,
            breaklinks = false, backref = false,
            colorlinks = true,
            linkcolor = blue, urlcolor = blue, citecolor = blue]
            {hyperref}
\usepackage{bm}
\usepackage{upgreek}
\usepackage{braket}
\usepackage{graphicx}

\begin{document}
\title{Ionization-induced Long-lasting Orientation of Symmetric-top Molecules}

\author{Long Xu}
\thanks{These authors contributed equally to this work}
\affiliation{AMOS and Department of Chemical and Biological Physics, The Weizmann Institute of Science, Rehovot 7610001, Israel}

\author{Ilia Tutunnikov}
\thanks{These authors contributed equally to this work}
\affiliation{AMOS and Department of Chemical and Biological Physics, The Weizmann Institute of Science, Rehovot 7610001, Israel}
\author{Yehiam Prior}
\email{yehiam.prior@weizmann.ac.il}
\affiliation{AMOS and Department of Chemical and Biological Physics, The Weizmann Institute of Science, Rehovot 7610001, Israel}
\author{Ilya Sh. Averbukh}
\email{ilya.averbukh@weizmann.ac.il}
\affiliation{AMOS and Department of Chemical and Biological Physics, The Weizmann Institute of Science, Rehovot 7610001, Israel}
\begin{abstract}
    We theoretically consider the phenomenon of field-free long-lasting orientation of symmetric-top molecules ionized by two-color laser pulses. The anisotropic ionization produces a significant long-lasting orientation of the surviving neutral molecules. The degree of orientation increases with both the pulse intensity and, counterintuitively, with the rotational temperature. The orientation may be enhanced even further by using multiple delayed two-color pulses. The long-lasting orientation may be probed by even harmonic generation or by Coulomb-explosion-based methods. The effect may enable the study of relaxation processes in dense molecular gases, and may be useful for molecular guiding and trapping by inhomogeneous fields.
\end{abstract}
\maketitle

\emph{Introduction}.---Field-free oriented molecules are essential in many studies, such as ultrafast dynamic imaging, molecular tomography, and electron diffraction, to name just a few.  A much more comprehensive list of applications may be found in a recent review by Koch et al. \cite{Koch2019Quantum}. Naturally, for practical applications, a sizable degree of orientation is beneficial.
One of the tools for inducing the molecular orientation is a non-resonant two-color laser pulse consisting of the fundamental wave (FW) and its second harmonic (SH).
Using such fields, two different orientation mechanisms have been identified and studied theoretically and experimentally \cite{Kanai2001,De2009,Oda2010,Spanner2012Mechanisms,Frumker2012,Znakovskaya2014,Kraus2015Observation}.
The first orientation mechanism, which is dominant at low to moderate (non-ionizing) intensities, relies on the interaction of the external fields with the molecular hyperpolarizability, which results in asymmetric torques that orient the molecules along the polarization direction of the SH field \cite{Kanai2001,De2009,Oda2010,Lin2018All,Xu2021Three}.
At high (ionizing) intensities, the dominant orientation mechanism \cite{Spanner2012Mechanisms} is different --  probability of ionization depends on the molecular orientation with respect to the polarization direction of the asymmetric electric field of the two-color pulse  \cite{Spanner2012Mechanisms,Frumker2012,Znakovskaya2014}. As a result, immediately after the pulse, the angular distribution of the surviving neutral molecules is asymmetric and has a non-zero orientation on average. Note that for linear molecules, this orientation disappears shortly after the excitation, but it periodically reemerges due to the phenomenon of rotational quantum revivals \cite{Averbukh1989,Robinett2004}.

Here, we theoretically investigate the ionization-induced orientation of \emph{symmetric-top molecules} excited by intense two-color femtosecond laser pulses. We demonstrate that in addition to the transient post-pulse orientation, and unlike linear molecules, there also exists a significant long-lasting orientation in these molecules. Long-lasting means that the orientation exists not only at the revival times but between the revivals too. In other words, the orientation signal has a non-zero baseline. Within the idealized model of non-interacting rigid rotors used here, this orientation lasts indefinitely. In practice, however, it will eventually be suppressed by additional physical effects, e.g., by intermolecular collisions in gas cell experiments.
Related effects of long-lasting orientation have been recently investigated in chiral \cite{Milner2019Controlled,Tutunnikov2020Observation,TutunnikovXu2020,Xu2021Three} and other non-linear \cite{Xu2020,xu2021longlasting,Xu2021Three} molecules excited by non-ionizing THz and laser pulses.

In what follows, we present our numerical analysis, outline our results on significant long-lasting orientation, and discuss its dependence on intensity and temperature. We conclude with a discussion of the experimental feasibility of observing the predicted effect.

\emph{Numerical methods}.---In our analysis, we simulate the rotational dynamics of symmetric-top molecules within the rigid rotor approximation both classically and quantum-mechanically. The symmetric-top molecules are excited by a  two-color laser pulse, consisting of the co-linearly polarized and phase-locked FW field and its SH.
The electric field is described by
\begin{equation}
    \mathbf{E}(t) = E_0 f(t) [\cos(\omega t)+\varepsilon\cos(2\omega t+\phi_0)] \mathbf{e}_Z,
    \label{eq:electric-field}
\end{equation}
where $E_0$ and $\omega$ are the peak amplitude and the carrier frequency of the FW, respectively. The laser pulse envelope is defined by $f(t)=\exp[-2\ln 2\,(t^2/\sigma^2)]$,
where $\sigma$ is the full width at half maximum (FWHM) of the pulse intensity profile, and $\mathbf{e}_Z$ is a unit vector along the laboratory $Z$ axis. Here, we set
$\varepsilon=1$ and $\phi_0=0$.

The Hamiltonian describing molecular rotation driven by a two-color laser pulse is
given by $H(t) =H_{r}+H_\mathrm{int}$, where $H_{r}$ is the rotational kinetic energy Hamiltonian and the interaction Hamiltonian is given by
\begin{equation}
    \label{eq:interaction-Hamiltonian}
    H_\mathrm{int}= V_\mathrm{pol}+V_\mathrm{hyp}+V_\mathrm{ion}.
\end{equation}
The field-polarizability and field-hyperpolarizability interaction terms are defined as
\cite{Buckingham2007Permanent}
\begin{equation}\label{eq:potential}
V_\mathrm{pol} = -\frac{1}{2} \sum_{i, j} \alpha_{i j} E_{i} E_{j},
V_\mathrm{hyp} = -\frac{1}{6} \sum_{i, j, k} \beta_{i j k} E_{i} E_{j} E_{k},\!\!
\end{equation}
where $E_{i}$, $\alpha_{i j}$, and $\beta_{ijk}$ are the components of the electric field vector, polarizability tensor, and hyperpolarizability tensor, respectively.
\begin{figure}[!t]
    \centering{}
    \includegraphics[width=\linewidth]{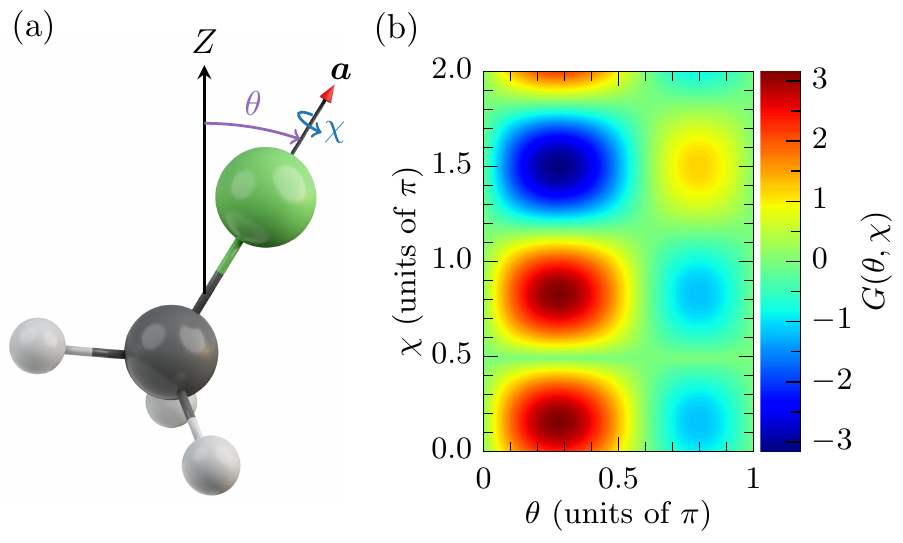}
    \caption{(a) $\mathrm{CH_3F}$ molecule.  Atoms are color-coded: black,
             carbon; gray, hydrogen; green, fluorine. $\theta$ is the angle
             between the molecular $a$ axis and the laboratory $Z$ axis,
             and $\chi$ represents the rotation angle about the $a$ axis.
             (b) Structure factor $G(\theta,\chi)$ [see Eqs. \eqref{eq:structure_factor} and \eqref{eq:structure_factor_1}]
             determining the angle dependence of the ionization rate.
    \label{fig:molecule}}
\end{figure}
The ionization depletion term $V_\mathrm{ion}$ is sensitive to molecular orientation.
For linear molecules, e.g., $\mathrm{HCl}$ \cite{Akagi2009Laser} and $\mathrm{CO}$ \cite{Li2011Orientation,Wu2012Multiorbital,Spanner2012Mechanisms}, the ionization rate depends on the angle between the molecular axis and the polarization direction.
For symmetric-top molecules, e.g., $\mathrm{CH_3F}$ and $\mathrm{CH_3Br}$, belonging to the $C_{3v}$ point group, the ionization rate depends on two angles, $\theta$ and $\chi$ (see Fig. \ref{fig:molecule}) \cite{Kraus2015Observation}.

In this work we consider $\mathrm{CH_3F}$ as our example symmetric-top molecule. The ionization process is modeled using a complex absorbing potential
\begin{equation}
    \label{eq:ionization}
    V_\mathrm{ion} = -\frac{i}{2} \Gamma(\theta,\chi,t),
\end{equation}
where the ionization rate $\Gamma(\theta,\chi,t)$ is defined as \cite{Kraus2015Observation}
(within the weak-field asymptotic theory)
\begin{equation}
    \label{eq:ionization-rate}
    \Gamma(\theta,\chi,t) =
    \begin{cases}
        W(t) |G(\theta,\chi)|^2, & E(t) > 0,\\
        W(t) |G(\pi-\theta,\pi+\chi)|^2, & E(t) < 0.
    \end{cases}
\end{equation}
Here, $E(t) = \mathbf{e}_Z \cdot \mathbf{E}(t)$, $W(t)$ is the field factor, and
$G(\theta,\chi)$ is the structure factor. The field factor is given by
\cite{Kraus2015Observation}
\begin{equation}
    W(t)= \frac{\kappa}{2}\left(\frac{4\kappa^2}{|E(t)|}\right)^{2/\kappa-1}
    \exp \left[ -\frac{2\kappa^3}{3|E(t)|} \right],
\end{equation}
where $\kappa=\sqrt{2I_p}$, and $I_p$ is the field-free energy of the highest occupied molecular orbital (HOMO).
We use the following model for the structure factor
\begin{equation}
    \label{eq:structure_factor}
    G(\theta,\chi) = \left[ \sin(\theta) + \frac{3}{2} \sin(2 \theta) \right] G_1(\chi),
\end{equation}
where
\begin{equation}
    \label{eq:structure_factor_1}
    G_1(\chi)=\begin{cases}
     \,\,\,\,\sqrt{1 + \sin(3 \chi)}, & 0\le\chi<7\pi/6,\\
     -\sqrt{1 + \sin(3 \chi)}, &7\pi/6\le\chi<11\pi/6,\\
       \,\,\,\,\sqrt{1 + \sin(3 \chi)}, & 11\pi/6\le\chi< 2\pi.
    \end{cases}\!\!\!
\end{equation}
$G(\theta,\chi)$ defined in Eqs. \eqref{eq:structure_factor} and \eqref{eq:structure_factor_1} closely approximates the structure factor of field-dressed HOMO of $\mathrm{CH_3F}$ with the largest dipole moment \cite{Kraus2015Observation} (the orbital from which the strong field ionization preferentially occurs).
The definition in Eq. \eqref{eq:ionization-rate} accounts for the oscillations of the laser electric field along the $Z$ axis.
We quantify the degree of orientation of surviving neutral molecules using the thermally averaged quantum expectation value of $\cos(\theta)$, $\braket{\cos(\theta)}$.
Further details on the quantum simulations can be found in \cite{xu2021longlasting,Tutunnikov2022Echo} (also see Sec. I of the Supplemental Material \cite{SupplementalMaterial}).

In classical simulations, we use the Monte Carlo approach to simulate the behavior of a classical ensemble consisting of $N=10^{7}$ sample molecules. A detailed description
can be found in \cite{Tutunnikov2019Laser,xu2021longlasting} (also see Sec. I of the Supplemental Material \cite{SupplementalMaterial}). Following the ionization depletion,
the classical degree of orientation is given by
\begin{align}
    \braket{\cos(\theta)}(t) =
    \sum\limits _{n=1}^{N}\rho(\theta_n,\chi_n,t)
    \cos(\theta_{n}),
    \label{eq:classical-orientation}
\end{align}
where the relative weight (non-ionized fraction) of the $n$-th molecule is
\begin{align}
    \rho(\theta_n,\chi_n,t)=
     N_\mathrm{neu}^{-1}\exp\left[-\int_0^t \Gamma(\theta_n,\chi_n,t^\prime)\,dt^\prime\right],
\end{align}
and the total number of surviving neutral molecules is
\begin{align}
    N_\mathrm{neu}=\sum\limits _{n=1}^{N}
    \exp\left[-\int_0^t \Gamma(\theta_n,\chi_n,t^\prime)\,dt^\prime\right],
\end{align}
Here, $\theta_n$ and $\chi_n$ are the time-dependent angles of the $n$-th molecule. The population of surviving neutral molecules is defined as $N_\mathrm{neu}/N$.

\begin{figure}[!t]
    \centering{}
    \includegraphics[width=\linewidth]{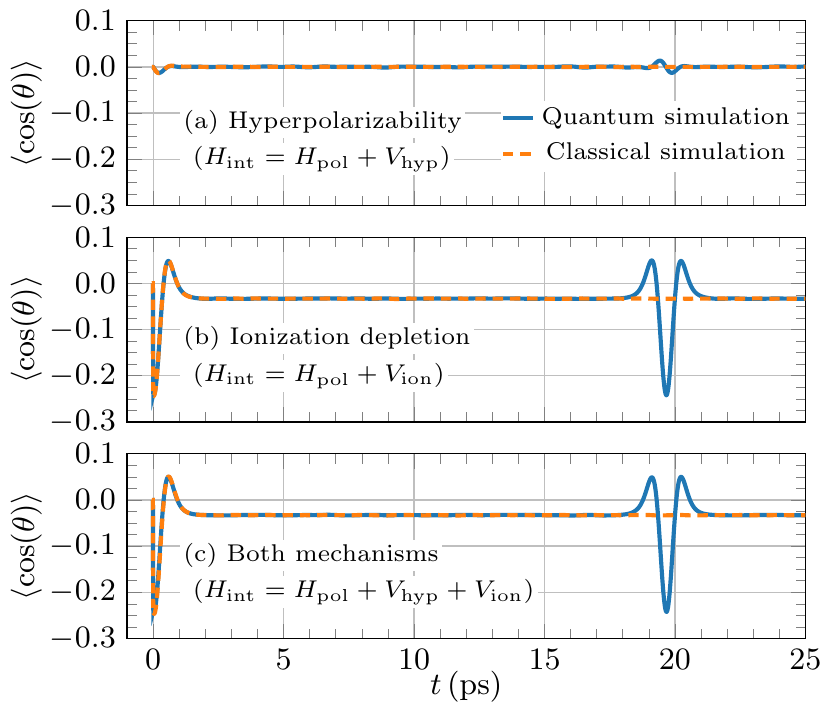}
    \caption{Time-dependent orientation factor at $T=300\,\mathrm{K}$ for different orientation mechanisms
             calculated classically and quantum-mechanically.
             Here the field intensity is $I_0=7\times 10^{13}\,\mathrm{W/cm^2}$ and the pulse duration is $\sigma=20\,\mathrm{fs}$.
             About 40\% of neutral molecules survive after the ionization.
     \label{fig:time-dependent}}
\end{figure}

\emph{Results}.---Molecular parameters of $\mathrm{CH_3F}$ are provided in Sec. II of the Supplemental Material \cite{SupplementalMaterial}.
Figure \ref{fig:time-dependent} shows the calculated, classically and quantum-mechanically, time-dependent orientation factor of neutral molecules following a single two-color pulse applied at $t=0$.
The parameters used in this calculation are: the rotational temperature is $T=300\,\mathrm{K}$, the laser wavelengths are 800\,nm (FW) and 400\,nm (SH), the peak intensity is $7\times 10^{13}\,\mathrm{W/cm^2}$, and $\sigma=20\,\mathrm{fs}$, see Eq. \eqref{eq:electric-field}.

The three panels of Fig. \ref{fig:time-dependent} show the orientation factor obtained for various combinations of interaction terms [see Eq. \eqref{eq:interaction-Hamiltonian}].
All cases include the field-polarizability interaction, $V_\mathrm{pol}\propto \cos^2(\theta)$ which is a symmetric function of $\theta$ (about $\theta=\pi/2$).
A torque-kick by such a potential results in molecular alignment only (for review, see \cite{StapelfeldtSeidman2003}). The two other terms, $V_\mathrm{hyp}$ and $V_\mathrm{ion}$ are asymmetric functions of $\theta$ and thus induce molecular orientation.
All three panels depict the immediate response of $\braket{\cos(\theta)}$ to the laser excitation near $t=0$.
This transient orientation effect is similar to the one studied in linear molecules excited by two-color laser pulses \cite{De2009,Oda2010,JW2010,Mun2018,Mun2019,Mun2020,MelladoAlcedo2020,Wang2020}.
At room temperature and field parameters used here, the transient molecular orientation induced by the field-hyperpolarizability interaction alone [Fig. \ref{fig:time-dependent}(a)] is negligible compared to the orientation resulting from the ionization depletion [Fig. \ref{fig:time-dependent}(b)].
Accordingly, the curves in Fig. \ref{fig:time-dependent}(b) and Fig. \ref{fig:time-dependent}(c) [in which all the interaction terms are included, see Eq. \eqref{eq:interaction-Hamiltonian}] are almost indistinguishable.
These results are consistent with previous results reported for linear molecules \cite{Spanner2012Mechanisms,Znakovskaya2014}, where it was shown that at high (ionizing) intensities, the orientation mechanism of ionization depletion dominates.

In this work, we focus on the long-term orientation existing in symmetric-top molecules. Figures \ref{fig:time-dependent}(b) and \ref{fig:time-dependent}(c) show that under the stated conditions and for these molecules, following the transient orientation the degree of orientation doesn't return to zero but persists at a constant value till the first revival and beyond.
\emph{This effect of ionization-induced long-lasting orientation doesn't exist in linear molecules, and it is the main result of this Letter}.
Note that the quantum and classical results agree well on the short time scale, during the initial transient response, and predict the same degree of long-lasting orientation on the long time scale, suggesting that the long-lasting orientation stems from a classical origin.
Long-lasting orientation has been previously observed in chiral \cite{Milner2019Controlled,Tutunnikov2020Observation,TutunnikovXu2020,Xu2021Three} and studied in other non-linear molecules \cite{Xu2020,xu2021longlasting,Xu2021Three} excited by non-ionizing THz and laser pulses.
In these cases, the orientation mechanisms, including the interactions with the polarizability and hyperpolarizability, were considered.
Here, the ionization depletion mechanism of ionization depletion gives rise to unprecedented degrees of long-lasting orientation at room temperature.
The degree of long-lasting orientation ($\approx-0.033$), as shown in Figs. \ref{fig:time-dependent}(b) and \ref{fig:time-dependent}(c), is an order of magnitude higher than values reported in previous studies.

\emph{Mechanism}.---Next, we discuss the mechanism behind this large  ionization-induced long-lasting orientation.
Under field-free condition, the symmetry axis (dipole) of symmetric-top molecules precesses around the (conserved, space-fixed) vector of angular momentum, whereas linear molecules rotate in a plane perpendicular to the angular momentum \cite{LANDAU}.
It is this  precession that is the source of the long-lasting orientation in symmetric-top molecules.
The degree of long-lasting orientation (with respect to the $Z$ axis) of a single symmetric-top molecule is given by the combination of three quantities $L_a L_Z/L^2$ \cite{Xu2020, xu2021longlasting} (see also Sec. III of the Supplemental Material \cite{SupplementalMaterial}), where $L_a$ and $L_Z$ are the projections of the angular momentum along the molecular symmetry axis [the molecular $a$ axis in Fig. \ref{fig:molecule}(a)] and the laboratory $Z$ axis, respectively, and $L$ is the magnitude of the angular momentum. Note that in the presence of $Z$-polarized pulses, $L_a$ and $L_Z$ are conserved quantities.

Initially, before the laser pulse, the molecules are isotropically distributed in space, namely there is an equal number of molecules having positive (along the $Z$ axis) and negative (against the $Z$ axis) long-lasting orientations.
Therefore, the ensemble-averaged long-lasting orientation $\braket{L_a L_Z/L^2}$ vanishes \cite{xu2021longlasting}.
The co-linearly polarized two-color laser pulse preferentially ionizes molecules oriented more or less along its polarization axis (with $\theta<\pi/2$) [see Fig. \ref{fig:molecule}(b)]. Thus, this selective ionization depletion breaks the symmetry of the molecular ensemble, generating a  non-zero long-lasting orientation of the surviving (not ionized) neutral molecules, as shown in Figs. \ref{fig:time-dependent}(b) and \ref{fig:time-dependent}(c).

For symmetric-top molecules, within the model used here, the long-lasting orientation is permanent.
In experiments, however, this orientation will be gradually destroyed by other physical effects such as intermolecular collisions, centrifugal distortion, and radiation emission caused by rapidly rotating molecular permanent dipole moments.
Furthermore, while the centrifugal distortion is known to lead to the decay of the revivals’ peaks due to the dephasing of the rotational states \cite{Damari2016Rotational,Babilotte2016Observation}, the average dipole remains almost unchanged (see \cite{Xu2020}). The radiative emission gradually decreases the rotational energy \cite{Damari2016Rotational,Babilotte2016Observation}, but the relative energy loss during a single revival is negligible for a rarefied molecular gas.

\begin{figure}[!t]
    \centering{}
    \includegraphics[width=\linewidth]{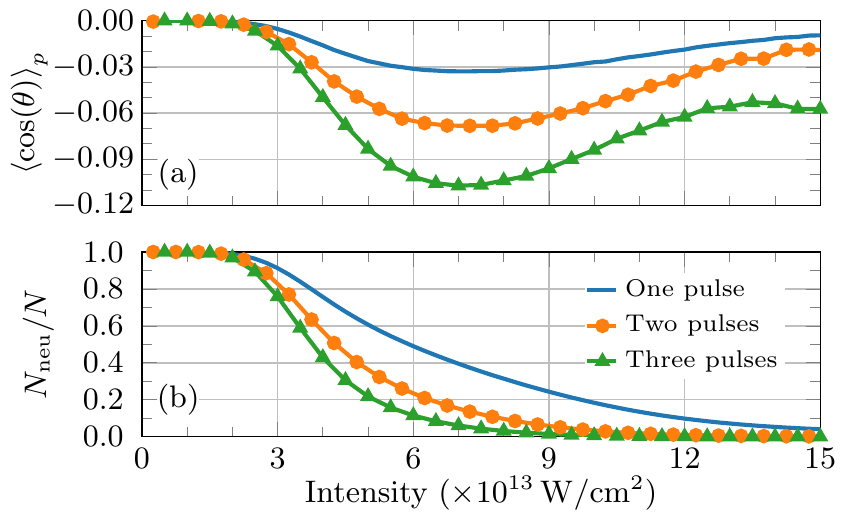}
    \caption{Classically calculated (a) long-lasting orientation and (b) population
             of neutral molecules after the pulse(s) as functions of the laser intensity.
             The time delay between pulses is 0.5\,ps. Here $T=300\,\mathrm{K}$ and
             $\sigma=20\, \mathrm{fs}$. Both orientation mechanisms are taken into account.
    \label{fig:intensity}}
\end{figure}

\emph{Intensity dependence}.---One of the ways to enhance the degree of the ionization-induced long-lasting orientation is by increasing the laser pulse energy --  peak intensity and/or pulse duration, or investing the higher energy in several pulses.
Figure \ref{fig:intensity} shows the classically calculated long-lasting  orientation factor, $\braket{\cos(\theta)}_p$, and the population of surviving neutral molecules as functions of the laser
intensity for single or multiple delayed pulses.
The value of $\braket{\cos(\theta)}_p$ is taken after the pulse(s), when the orientation factor reaches a constant value.
As expected, with the increasing input energy, the long-lasting orientation factor (in absolute value) initially grows [for $I_0<7\times 10^{13}\,\mathrm{W/cm^2}$, see Fig. \ref{fig:intensity}(a)], while the population of surviving neutral molecules decreases monotonically [see Fig. \ref{fig:intensity}(b)].

After reaching its maximum value, the long-lasting orientation decreases with increasing laser intensity.
The reason for this decrease is the structure factor $G(\theta,\chi)$ shown in Fig. \ref{fig:molecule}(b).
Due to the anisotropy of the ionization process, molecules at $\theta \approx 0,\,0.6\pi,\,\pi$ have a relatively low ionization rate. Therefore, when the ionization saturates at high laser intensities, only these molecules survive, but these molecules have a relatively low contribution to the long-lasting orientation. The combination of increased ionization yield, but counterproductive selectivity of low-contributing molecules limits the usefulness of increasing the laser intensity beyond a certain point.

A way to avoid the limits imposed on high intensities is to use multiple pulses instead of a single strong pulse.  Consider the application of several delayed two-color pulses. Figure \ref{fig:intensity}(a) shows that the maximum long-lasting orientation is about $-0.033$ for one pulse, $-0.069$ for two pulses, and $-0.107$ for three pulses. Here, the time delay between each pulse is set to 0.5\,ps, a relatively long time delay which allows the molecules to rotate between the pulses. This way, after each  pulse additional neutral molecules rotate to an orientation more favorable for ionization by the next pulse. This approach overcomes the ionization saturation limit that exists in the case of a single pulse, and is similar to the use of multiple pulses for achieving enhanced alignment while avoiding ionization \cite{Averbukh2003, Averbukh2004}. Naturally, adding more pulses results in a progressively lower population of surviving neutral molecules. Nevertheless, for a fixed population, a higher long-lasting orientation is achieved by applying multiple delayed pulses. For example, for $N_\mathrm{neu}/N\approx 40\%$, the long-lasting
orientation is about $-0.033$ for one pulse, $-0.049$ for two pulses, and $-0.054$ for three pulses.
\emph{This observation is an additional main result of this Letter.} Optimization of the time delay may allow further enhancement of the long-lasting orientation.

\begin{figure}[!t]
    \centering{}
    \includegraphics[width=\linewidth]{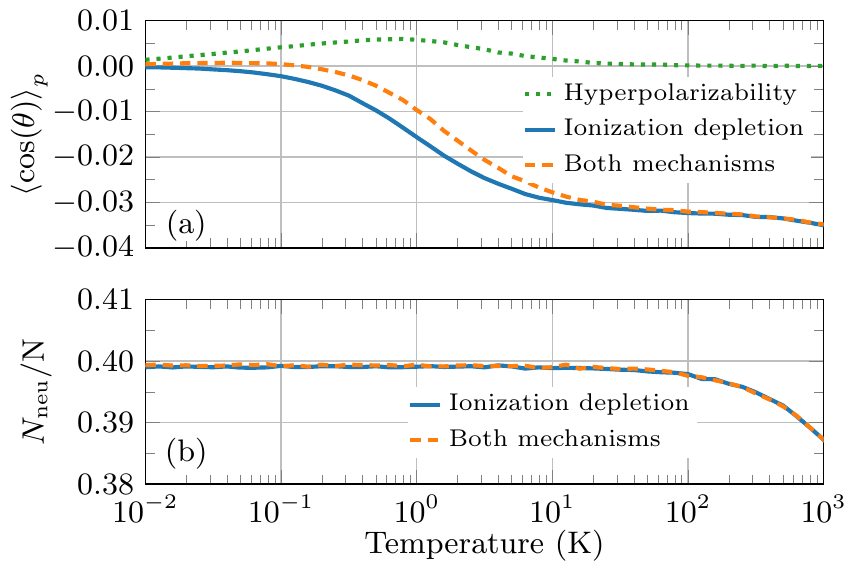}
    \caption{Classically calculated (a) long-lasting orientation factor and (b) population
             of neutral molecules as functions of temperature for different orientation mechanisms.
             The laser parameters are the same as in Fig. \ref{fig:time-dependent}:
             $I_0=7\times 10^{13}\,\mathrm{W/cm^2}$, $\sigma=20\,\mathrm{fs}$.
             The population for the orientation mechanism of hyperpolarizability  remains 1.
     \label{fig:temperature}}
\end{figure}

\emph{Temperature dependence}.---Next, we consider the temperature dependence of the long-lasting orientation as depicted in Fig. \ref{fig:temperature}. At $T=0$, the long-lasting orientation is zero for all three curves shown in Fig. \ref{fig:temperature}(a), as a result of $L_a=L_Z=0$ (these are conserved quantities).
There is an optimal temperature ($T<1\,\mathrm{K}$ in this example) at which the long-lasting orientation induced by
the field-hyperpolarizability interaction (together with the field-polarizability interaction) reaches the maximum ($\approx 0.006$). Above the optimal temperature, the long-lasting orientation decays with increasing temperature \cite{xu2021longlasting}. In sharp contrast, the ionization-induced long-lasting orientation factor (in absolute value) increases monotonically with the temperature, which is another principal finding of this work.
As mentioned above, the long-lasting orientation is given by $L_a L_Z/L_f^2$, where $L_f=|\mathbf{L}_f|$ is the magnitude of the angular momentum after the pulse.
$L_Z$ and $L_a$ are conserved in the case of excitation by linearly polarized laser pulses and thus are functions of temperature only. The short two-color pulse has a two-fold effect: (i) it (almost) instantaneously ionizes the molecules at particular angles (the effect described by $V_\mathrm{ion}$), and (ii) it changes the molecular angular momentum, $\mathbf{L}_f=\mathbf{L}_i+\delta \mathbf{L}$, where $\mathbf{L}_i$ is the initial angular momentum, and $\delta \mathbf{L}$ is the change caused by $V_\mathrm{pol}$.
Around the room temperature (in the examples considered here), the effect of $V_\mathrm{pol}$ becomes negligible, i.e., $L_a L_Z/L_f^2\approx L_a L_Z/L_i^2$, and the degree of long-lasting orientation reaches a constant value determined by $V_\mathrm{ion}$.
At lower temperatures, $V_\mathrm{pol}$ has a detrimental effect by increasing the total angular momentum, such that $L_a L_Z/L_f^2 < L_a L_Z/L_i^2$ effectively lowers the long-lasting orientation.

At temperatures above room temperature, the ionization can no longer be considered instantaneous because the molecules rotate fast enough to change their orientation during the pulse. Accordingly, more molecules get ionized, as seen from the populations in Fig. \ref{fig:temperature}(b).
The higher the ionization yield, the higher asymmetry of the molecular ensemble, manifesting in the higher long-lasting orientation.

\emph{Conclusions}.---We have theoretically demonstrated a sizable ionization-induced long-lasting orientation of symmetric-top molecules excited by two-color laser pulses.
The mechanism leading to this observation is the selective ionization of the polar molecules at particular angles with respect to the laser's electric field, and the  ability of symmetric-top molecules to precess around the fixed angular momentum vector.
We show that using a proper sequence of delayed two-color pulses allows for enhancing the long-lasting orientation of neutral molecules without drastic depletion of their population. Due to the required precession, the enhanced orientation favors high temperature (room temperature and above in our examples). The long-lasting orientation may be measured with the help of Coulomb explosion \cite{Znakovskaya2014}, or by observing second (or higher order) harmonic generation in the gas phase, which is sensitive to the lack of inversion symmetry \cite{Frumker2012Oriented,Frumker2012Probing,Kraus2012High}.
Long-lasting orientation can pave the way to the study of relaxation processes in dense molecular gases that are not  otherwise accessible to direct probing (see analogous applications of persistent alignment \cite{Hartmann2012Quantum,Vieillard2013Field}).
In addition, molecular focusing, guiding, and trapping by inhomogeneous fields rely on molecular orientation \cite{Gershnabel2011Electric,Gershnabel2011Deflection,Kupper2012},  and their observation may be facilitated by this long-lasting orientation effect.
\begin{acknowledgments}
    L.X. is a recipient of the Sir Charles Clore Postdoctoral Fellowship.
    This research was made possible in part by the historic generosity
    of the Harold Perlman Family.
\end{acknowledgments}

\end{document}